\documentclass[12pt]{article}
 
\usepackage{amssymb}
\usepackage{graphicx}
\usepackage{fourier}

\newcommand{\vs}{\vspace{15pt}} \newcommand{\n}{\noindent} 
\newcommand{\rf}[1]{(\ref{#1})}
\newcommand{\ba}{\begin{array}} \newcommand{\ea}{\end{array}}
\newcommand{\be}{\begin{equation}} 
\newcommand{\btb}{\begin{tabular}}\newcommand{\etb}{\end{tabular}}
\newcommand{\ee}[1]{\label{#1}\end{equation}}
\newcommand{\bi}{\bibitem}

\newcommand{\dss}{\displaystyle}
\newcommand{\bfl}{\begin{flushleft}}\newcommand{\efl}{\end{flushleft}}

\newcommand{\al}{\alpha} \newcommand{\bt}{\beta}\newcommand{\g}{\gamma}  
\newcommand{\de}{\delta}\newcommand{\De}{\Delta}

  \newcommand{\R}{\mathbb R}

\newcommand{\PCU}{{\cal P}_\uparrow}

\newcommand{\ry}{{\bf r}}

\newcommand{\bite}{\begin{itemize}} \newcommand{\eite}{\end{itemize}}

\begin{document} 
\title{On Spavieri's conundrum \\ [6pt] or the shadow of the twin paradox \\ [6pt]  \small A submission to the One-Way Linear Effect (OWLE) Award}
\author{Marco Mamone-Capria\\ \small Dipartimento di
Matematica - via Vanvitelli, 1 - 06123 Perugia - Italy \\ \small {\sl E-mail}:
\texttt{marco.mamonecapria@unipg.it}}
\normalsize \maketitle

\small
{\bf Abstract} We deal with a problem concerning a supposed inconsistency in the special relativistic treatment of the so-called linear Sagnac effect. It is shown that, under modern clothes, the root of the difficulty perceived by some authors lies in their uneasiness with the standard solution of the twin paradox. In particular, since the linear Sagnac effect is an absolute effect, no tinkering with conventionality of simultaneity, so far as it preserves the physical content of special relativity, would get us out of the supposed trouble. 

\normalsize

\tableofcontents

\section{Introduction}

Gianfranco Spavieri, a well-known Italo-Venezuelan physicist working in the foundations of relativity, has been advertising in a number of papers \cite{sp17, sp19, sp22} the claim, previously endorsed by the late Italian physicist Franco Selleri \cite{s10, s12}, that the so-called Linear Sagnac Effect (LSE) \cite{wzyl03, wzy04} cannot be reasonably interpreted within special relativity (SR).

Recently, he threw a challenge to the foundations of physics' community \cite{sp25}, based on a problem that will be quoted in the following section.

I shall cite, and refer to, the text of the announcement of this challenge \cite{sp25}, although it is essentially excerpted from the Appendix of \cite{sp22}.

I have discussed at length the Sagnac issue (both the original circular version and the more recent linear one) in \cite{mm22}, and I flattered myself that I had put to rest the controversy. This paper is meant as a supplement to the previous paper.

Among the added material here I consider the case in which the emission event occurs at an arbitrary position of the clock (or, as I prefer to call it, the clock-emitter-detector, or {\sl ced}). I also discuss the crucial and clarifying relationship of the linear Sagnac effect with the twin paradox, proving that the perceived `time gap' in the LSE and the differential aging (which is a genuine consequence of SR, of course, that no transfer to a nonstandard simultaneity can ever abolish) have the same source.

\section{The Conundrum}

Here is the text of the award challenge (from now on `the Challenge'), including the following Fig.1:

\begin{figure}[htp]
\centering
\includegraphics[totalheight=0.4\textheight]{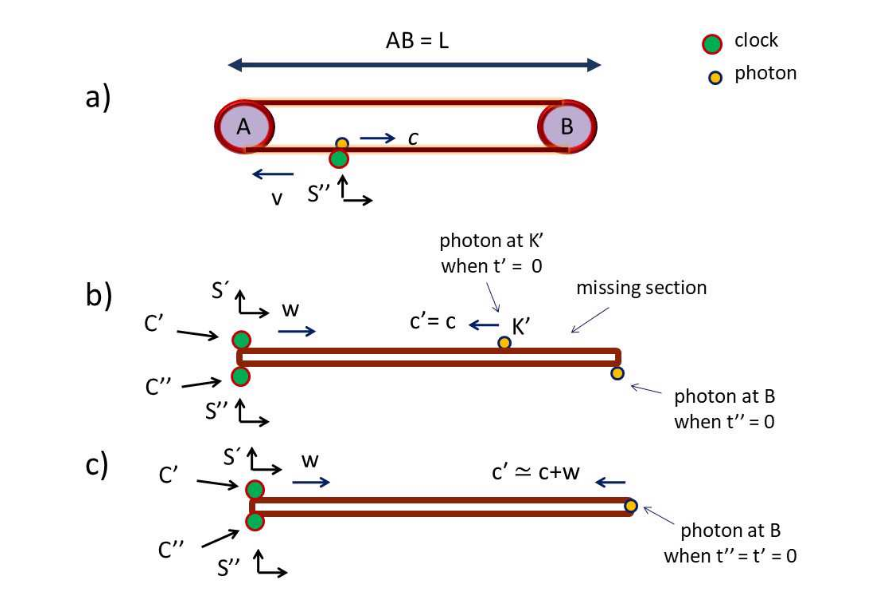}
\caption{Linear Sagnac effect}
\end{figure}

\begin{quote}\small
{\bf The exercise}. In the linear Sagnac effect of Fig. 1, an optical fibre (with refractive index $n=1$) slides at speed $v$ on the two pulleys A and B with AB $=L$. The measuring device C$^\ast$ (a clock) is fixed to the fiber and may move with it from the lower to the upper section while an emitted counter-propagating photon performs the round trip in the invariant time interval $T$. For the counter-moving photon, the round-trip interval $T$ measured by C$^\ast$ co-moving with the fiber ($\g^{-2} = (1+ v/c)(1-v/c) = 1- v^2/c^2$) is (Post 1967),

\be T = \frac{2L}{\g (c+ v)} = \frac{2\g L}{\g^2  (c+ v)} = \frac{2\g L (1-v/c)}{c} \simeq \frac{2L (1-v/c)}{c}. \ee{LSE-time}

\n
which is independent of the initial position of C$^\ast$ and can be calculated even in Newtonian physics ($\g \simeq 1$).

The fiber length $\De L_{loc} = c_{loc} \De T_{loc}$ traversed by the photon along any fiber section, corresponds to the length determined using the local (differential) speed $c_{loc} = c$ and local clocks measuring the time interval $\De T_{loc}$ while at rest with the fiber in that section.

{\bf The challenge}.

{\bf Demonstrate the following}. Using the Lorentz transformation (LT) in the context of flat spacetime, show that, in the one-way linear effect discussed above, the photon, traveling at the {\sl local} speed $c$ along each section of the optical fiber, covers the {\sl whole} fiber length $\simeq 2L$ in the proper time interval $T$ measured by the clock C$^\ast$.
\end{quote}\normalsize

The text is accompanied by a number of remarks explaining what is viewed as an inconsistency: a ``time gap'' in the way special relativity (SR) deals with the LSE, and which would be repaired - it is assumed - by resorting to absolute simultaneity. 

We shall see that the supposed `time gap' arises from using a mistaken procedure to compute the relevant quantity, and that no inconsistency in the foundations of SR is involved. I will provide as painstaking and explicit a treatment as reasonable within the scope of this paper.

There are some preliminary general remarks, however, that deserve to be made. 

One issue is that in SR a ced not bound to a single Minkowskian coordinate system can only measure the {\sl proper time} between events occurring (for instance, emission or reception events) at its own (changing) place. Thus, if we want to measure the length of the LSE apparatus, the main legitimate method is to use an inertial coordinate system with which the axles of the pulleys are at rest: this would enable us to obtain the {\sl proper length}  of the apparatus. On the other hand, for an observer moving with the ced, the LSE apparatus simply does not exist as a spatial entity.

In general, in SR a length, or a distance, measured in a given coordinate system, and another length, or distance, measured in a coordinate system moving with respect to the former, cannot be added in a meaningful way.    

Related to this is the advice that when dealing with SR one should avoid as much as possibile using {\sl space} diagrams (like Fig.1 and my Fig.4). Instead one should employ systematically {\sl space-time} diagrams - even when doing Newtonian computations. This simple device is both clarifying and preventive of a number of misunderstandings, and I hope the following treatment will bring it home.

\section{Basic concepts and notation on Minkowski space-time}

To fix notation, let us recall what an $n$-dimensional ($n\geq 2$) Minkowski space-time is (for my notation and basic definitions see \cite{mm16}). It is an $n$-dimensional real affine space $M$ endowed with a Lorentzian metric $g$ on its associated vector space, and with both time and space orientations. 

Equivalently, it can be described as a pair $(M, \Phi)$, where $\Phi$ is a $\PCU^+$-orbit on the set of bijections $\mbox{Bi} (M, \R^n)$ and $\PCU^+(n)$ is the proper orthochronous Poincar\'e group on $\R^n$.

The elements of $M$ are called events, and the elements of $\Phi$ are Minkowski coordinate systems ({\sl cs}) If $\phi\in \Phi$ and $p\in M$, we shall write: 

\[\phi (p) = (\ry (p), t(p)) =  (x^1 (p), x^2 (p), x^3(p), t(p)),\] 

\n
or equivalently:

\[ p \stackrel{\phi}{\equiv} (\ry (p),t). \]

Needless to say, standard Minkowski space-time is 4-dimensional. However, to deal with 2-dimensional physical processes one can use, formally, 3-dimensional Minkowski space-time, and in the case of the (idealized) LSE one may omit two spatial dimensions altogether, and reduce to the 2-dimensional  Minkowski space-time.  

If $g$ is the standard Lorentz form on $\R^2$, we denote by $\hat{g}$ the associated quadratic form, that is: 

\[ \hat{g} (\xi, \eta) = (\xi)^2 - c^2 (\eta)^2. \]
\n
where $c$ is the speed of light in empty space (or in the aether). We shall use the same symbols $g$ and $\hat{g}$ also for the vector space associated to $M$.

A vector $w$ is spacelike if it is either 0 or such that $\hat{g} (w)>0$; otherwise, it is timelike (resp. lightlike) if $\hat{g}(w)<0$ or $\hat{g}(w)=0$. Timelike or lightlike vectors are called causal; they are future (resp. past) if they belong (resp. do not belong) to the connected component of the set of causal vectors which is selected by the time orientation.

The Lorentzian module of a vector $w$ is $|w| := |\hat{g}(w)|^{1/2}$. 

One important fact that will be referred to in the following is that if $p,q$ are two events such that $q-p$ if future timelike, there are (infinitely many) cs's $\phi \in \Phi$ such that $p$ and $q$ occupy the same {\sl place} (i.e. $\ry (p) = \ry(q)$), and in this case the time difference between $p$ and $q$ is the same for all such $\phi$'s:

\be \De t = t(q)- t(p) = \frac{1}{c} |q-p| \ee{proti}

This is the {\sl proper time} of the line segment $pq$, and the {\sl Proper Time Principle} \cite{mm01} extends both definition and physical interpretation to any timelike piecewise smooth motion (and to its image, or worldline) in space-time. 

For consistency with \cite{mm22}, I shall adopt the following notation and rules of translation from my treatment to \cite{sp25}:

\[ \bt: = v/c, \; \al: = \frac{1}{\sqrt{1-\bt^2}} \leadsto \g , \; \ell \leadsto L. \]

\section{The Linear Sagnac Effect, classically}

We denote by $\phi$ a fixed Minkowskian coordinate system (cs) $(x,t)$ such that the apparatus of the LSE is at rest, with the left pulley $A$  having $x=0$ for all $t$.  This is our laboratory cs. The length of the apparatus with respect to $\phi$, that is, its rest length or proper length,  is denoted by $\ell$.

When expressing the coordinates of an event $p$ in this (and only this) cs $\phi$ we shall write simply $p \equiv (x,t)$, omitting the superscript.

In the apparatus we have an optical fiber with refractive index $n=1$ rolling counter-clockwise. This can be represented in a 2-dimensional Minkowski space-time by assuming the idealization that the radius of each pulley is negligible with respect to $\ell$, and thus formally equating $A$ and $B$ to mirrors, both for the photons and for the ced.

The ced moves together with the fiber at a constant speed, except at the turning point; every time the right pulley $B$ (resp. the left pulley $A$), is reached by the ced, its speed is converted from $v (> 0)$ into $-v$ (resp. from $-v$ to $v$). 

Let $p^\ast$ be the event at time $t=0$ of emission of a photon in the direction opposed to the motion of the fiber (a {\sl counter-propagating photon}), $p^-$ the reflection event at $A$, and $r$ the turning event of the ced at $B$. 

\n
N.B. {\sl In the literature \cite{wzyl03} these photons, or beams, are called ``counter-propagating'' (that is, one travelling clockwise, and the other counterclockwise), while in the present article, to conform to the text of the Challenge, we are calling `` counter-propagating'' the photon moving opposite to the motion of the fiber, the other being called ``co-propagating''}.

Clearly:

\be \ba{rcl} p^\ast &\equiv & (x^\ast, 0), \\ p^- &\equiv & (0, x^\ast/c), \\ r &\equiv & (\ell, \frac{\ell - x^\ast}{v}), \ea \ee{cophi}

\n
with $0<x^\ast<\ell$.

There are two possibilities, according to whether the return event  $q^-$ of the photon to the ced occurs in the lower or in the upper section of the apparatus, that is,  before or after the ced has reached $B$. One can use the formalism described in the previous section to show that this depends on whether $r - p^-$ is causal (i.e. timelike or lightlike) or spacelike. 

Explicitly, since

\[ \hat{g} (r-p^-) = \hat{g} (\ell, \frac{\ell - x^\ast}{v} - \frac{x^\ast}{c}) = \ell^2 - c^2 \left(\frac{\ell- x^\ast}{v} - \frac{x^\ast}{c}\right)^2 , \]

\n
after a few computations one gets that the photon reaches the ced in the upper section if and only if:

\be x^\ast > \frac{1-\bt}{1+\bt} \ell =: \hat{x}, \ee{divide-1}

\n
or, equivalently:

\be v > \frac{\ell - x^\ast}{\ell + x^\ast} c. \ee{divide-2}

\n
For instance, if we assume $x^\ast$ to be the middle point of the lower section of the apparatus, then it follows that $v > c/3$, which is obviously unrealistic in a LSE experiment (as I stated in \cite{mm22}); in order for a technically feasible rolling speed to be sufficient, the emission place will have to be (very) close to $\ell$. 

For instance, in the experiments described in \cite{wzy04} ``the length of the fiber loop is 4.1 m and the speeds of the motion are from 0.00191 to 0.211 m/s''. This corresponds to 

\[ \bt = 6.36 \times 10^{-12}  \div  7.03 \times 10^{-10}, \]

\n
that is,

\[ \hat{x} = 2.05 \cdot (0,99999999998728 \div 0,99999999859399)\; \mbox{m}.\]

For the computations in the next subsections it is useful to parametrize the ced's worldline (that is, speaking in terms of $\phi$'s space, from $x^\ast$ to $\ell$ and then back to 0):

\be \phi \circ \g (t) =\left\{ \ba{rcl} (x^\ast + vt, t) \; &\mbox{if}& \; t\in [0, (\ell- x^\ast)/v]
\\  (\ell - v(t - (\ell-x^\ast)/v), t) \; &\mbox{if}& \; t\in [(\ell- x^\ast)/v, (2\ell- x^\ast)/v]
\ea\right. \ee{param}

We shall now consider the two cases in sequence.

\subsection{The counter-propagating photon reaches the ced in the lower section}

\begin{figure}[htp]
\centering
\includegraphics[totalheight=0.4\textheight]{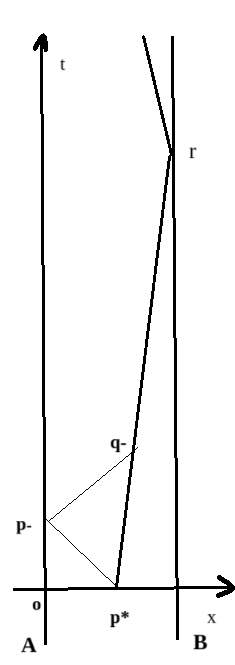}
\caption{Return occurring in the lower section}
\end{figure}

\n
We assume that \rf{divide-1} does {\sl not} hold.  Then the reception event $q^-$ is identified by the equality:

\[ q^- \equiv (0, x^\ast/c) + s (c,1) = (x^\ast + vt (q^-), t(q^-)), \]

\n
which implies

\[ sc = x^\ast + vt(q^-), \; s = t(q^-),\] 

\n
and therefore

\be t(q^-) = \frac{2x^\ast}{c(1-\bt)}. \ee{lower}

\n
This is the time from emission to reception without \rf{divide-1}, and it depends on $x^\ast$: 

\be \De t (p^\ast, q^-) = \frac{2x^\ast}{c(1-\bt)}.  \ee{timelab-0}

Moreover: 

\be q^- \equiv \left(\frac{1+\bt}{1-\bt} x^\ast ,  \frac{2x^\ast}{c(1-\bt)}.\right) \ee{qu-meno-lower}

\n
and the times of percurrence before and after $p^-$ (in the single lower section) are:

\[ \ba{rcl} \De t (p^\ast, p^-) &=& \dss \frac{x^\ast}{c}
\\ [6pt] \De t (p^-, q^-) &=&  \dss\frac{2x^\ast}{c(1-\bt)} - \frac{x^\ast}{c}, \ea\]

\n
with the former always lesser than the latter.

\subsection{The counter-propagating photon reaches the ced in the upper section}

\begin{figure}[htp]
\centering
\includegraphics[totalheight=0.35\textheight]{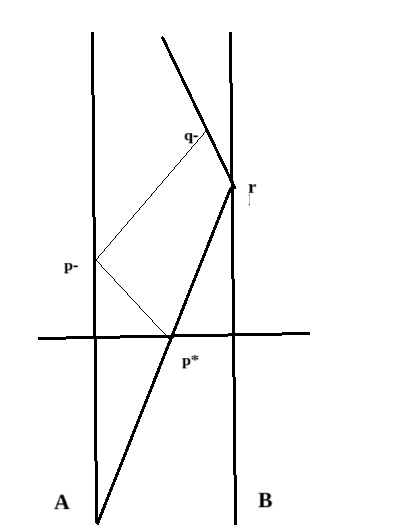}
\caption{Return occurring in the upper section}
\end{figure}

\n
In this case we have to use the second condition in \rf{param}, and require: 

\[ q^- \equiv (0, x^\ast/c) + s (c,1) = (\ell - v(t - (\ell-x^\ast)/v), t). \]

\n
By the same procedure as above we obtain:

\be t(q^-) = \frac{2\ell}{c (1+\bt)}. \ee{upper} 

\n
This is the time from emission to reception under inequality \rf{divide-1}, and it does not depend on $x^\ast$ (as correctly stated in \cite{sp25}). So we can also write:

\be \De t (p^\ast, q^-) = \frac{2\ell}{c (1+\bt)}. \ee{timelab}

Moreover:

\be q^- \equiv (\frac{2\ell}{1+\bt} - x^\ast, \frac{2\ell}{c (1+\bt)}). \ee{qu-meno}

Instead, the times of percurrence of the photon in the two sections of the apparatus are:

\be\ba{rcl}  \De t (p^\ast, p^-) &=& \dss\frac{x^\ast}{c}  \;\; (\mbox{lower section})
\\ [6pt] \De t (p^-, q^-) &=&  \dss \frac{2\ell}{c (1+\bt)} - \frac{x^\ast}{c} \;\; (\mbox{upper section}). \ea \ee{times-II} 

Notice that, taking into account \rf{divide-1}, we obtain that the following conditions are equivalent:

\be \ba{rcl} t(r) - t(p^\ast) &\geq& t(q^-) - t(r)  
\\ [6pt] t(q^-) - t(p^-) &\geq& t(p^-) - t(p^\ast) 
\\ [6pt] \dss \frac{1-\bt}{1+\bt}\ell \leq &x^\ast& \leq \dss\frac{\ell}{1+\bt}\ea\ee{diffty} 

\n
In particular, the only value of $x^\ast$ for which the two time differences, both in \rf{diffty}$_1$ and in \rf{diffty}$_2$, coincide is $x^\ast = \ell /(1+\bt)$.

\subsection{The linear Sagnac effect}

\begin{figure}[htp]
\centering
\includegraphics[totalheight=0.4\textheight]{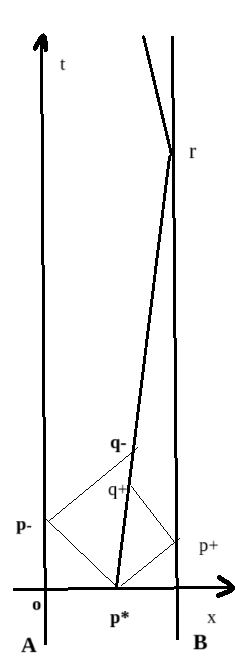}
\caption{LSE with return in the lower section}
\end{figure}

The linear Sagnac effect is the interferometric detection of a fringe shift due to the delay in reception of two simultaneously emitted photons by the ced along the optical fiber in opposite directions. 

A special case of the LSE, the one with $x^\ast$ chosen in order to more closely model the standard Sagnac effect, has been discussed in \cite{mm22}. Here we shall give the general formulas of the LSE according to whether the counter-propagating photon is received by the ced in the lower or in the upper section of the fiber. For the following computations we need to add to \rf{cophi} the following:

\be p^+ \equiv (\ell, \frac{\ell - x^\ast}{c}). \ee{cophi-1}
      
The reception event of the co-propagating photon is identified by:

\[ q^+ \equiv (\ell, \frac{\ell - x^\ast}{c}) + s (-c,1) = (x^\ast + v t(q^+), t(q^+)), \] 

\n
and following the same procedure as for $q^-$ we obtain

\be t(q^+) = \frac{2(\ell- x^\ast)}{c(1+\bt)}. \ee{qu-piu}
      
Notice that, whether the counter-propagating  photon is received in the upper or in the lower section, we have that the co-propagating photon is always received in the lower section, that is $t(q^+) < t(r)$ (a consequence of $\bt <1$). 

Thus we can immediately compute in the second case, by using \rf{qu-meno}:

\be (\De t)_{II} t (q^+, q^-) = \frac{2 x^\ast}{c(1+\bt)}. \ee{sa-upper}

\begin{figure}[htp]
\centering
\includegraphics[totalheight=0.4\textheight]{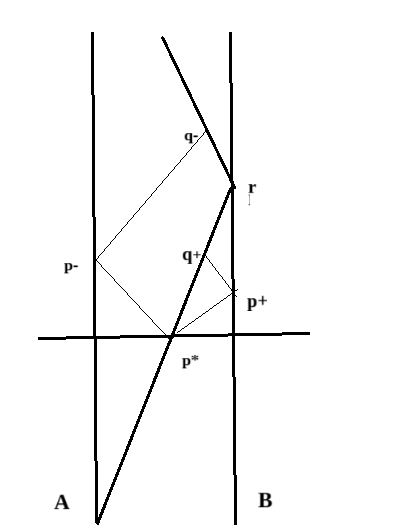}
\caption{LSE with return in the upper section}
\end{figure}

For the first case it remains to compute $t(q^-)$:

\[ q^- \equiv (0, \frac{x^\ast}{c}) + s(c,1) = (x^\ast + vt, t),\]

\n
which gives

\[ t (q^-)_I = \frac{2x^\ast}{c(1-\bt)}, \]

\n
and therefore:

\be (\De t)_I (q^+, q^-) = \frac{2(2x^\ast - \ell (1-\bt))}{c (1-\bt^2)}. \ee{sa-lower}. 

Thus the measured delay depends on $x^\ast = ]0,\ell[$ according to the continuous function:

\be (\De t) (x^\ast) = \left\{\ba{rcl} \frac{2(2x^\ast - \ell (1-\bt))}{c (1-\bt^2)} \; &\mbox{if}&  x^\ast \leq \hat{x}
\\ [6pt]   \frac{2 x^\ast}{c(1+\bt)} \; &\mbox{if}&  x^\ast \geq \hat{x} \ea\right. . \ee{LSE}

Notice that in the fist sub-interval $]0, \hat{x}[$ this function is negative when 

\[ x^\ast \leq \frac{\ell (1-\bt)}{2}, \]

\n
corresponding to a change in the shift of the fringes.

So far our treatment has been neutral between SR and classical physics, and could be endorsed by anyone rejecting SR, and accepting, instead, that 1) $\phi$ is an aether cs, 2) only absolute simultaneity is physically viable.

\section{Proper times and length contraction}

What does SR add to the previous treatment? There are two features. 

First, it predicts, from the results so far obtained, not just the time as measured in $\phi$, but the proper time between events in the ced's worldline {\sl as measured by the ced itself}. This distinction would be vacuous, of course, in classical physics: no other `proper time' exists except `absolute time' (in our case, $\phi$'s time), insofar as it is assumed that the ced is operating correctly.

In the LSE setting, if we set the ced's clock so that the proper time $\tau$ has its origin at $p^\ast$ (that is, $\tau (p^\ast) = 0$), then the proper time from $p^\ast$ to $q^-$ is obtained by factoring out the time dilation from  the time differences in $\phi$; we have the same denominator in the two terms, since $\al$ remains the same when $v$ is substituted by $-v$:

\[ \ba{rcl} \De \tau = \tau (q^-) - \tau (p^\ast) &=& (\tau (q^-) - \tau (r)) + (\tau (r) - \tau (p^\ast)) 
\\ [8pt] &=& \dss\frac{t (q^-) - t (r)}{\al} + \frac{t (r) - t (p^\ast)}{\al}
\\ [8pt] &=& \dss\frac{t (q^-) - t (p^\ast)}{\al} 
\\ [8pt] &=& t(q^-) \sqrt{1-\bt^2}. \ea\]

\n
Therefore, in the first case (\S 4.1):

\be \De \tau_I = \frac{2x^\ast\sqrt{1-\bt^2}}{c(1-\bt)}, \ee{proper-lower} 

\n
and in the second case (\S 4.2):

\be \De \tau_{II}  = \frac{2\ell\sqrt{1-\bt^2}}{c (1+\bt)}. \ee{proper-upper} 

These formulas can also be derived directly from \rf{timelab-0} and \rf{timelab} by simply using the formula for the time dilatation.

The second feature introduced by SR is length contraction.

Since the optical fiber is linked to the two $\phi$-places $A$ and $B$, the length of each of its two sections according to $\phi$ is just $\ell$. 

This implies that, according to the cs $\phi''$ (resp. $\phi'$) comoving with the lower (resp. upper) section of the fiber, the proper length of the corresponding section must be:

\be \ell_0 = \frac{\ell}{\sqrt{1-\bt^2}}. \ee{properlength}

\n
This formula will be essential in the solution of the Challenge. 

As to the LSE, there is no doubt that SR predicts it as well. In fact, by taking into account the time dilatation as above, we obtain the continuous function:

\be (\De \tau) (x^\ast) = \left\{\ba{rcl} \frac{2(2x^\ast - \ell (1-\bt))}{c \sqrt{1-\bt^2}} \; &\mbox{if}&  x^\ast \leq \hat{x}
\\ [6pt]   \frac{2 x^\ast \sqrt{1-\bt^2}}{c(1+\bt)} \; &\mbox{if}&  x^\ast \geq \hat{x} \ea\right. . \ee{LSE-rel}

This shows that the argument presented in the Challenge is not relevant to the issue of whether SR predicts the LSE effect: clearly it does predict it.





\n 
N.B. {\sl From now on we shall discuss the second case only, and `the photon' will always mean `the counter-propagating photon'}.

\section{Does the photon traverse the whole fiber?}

At the core of the Challenge there is the following issue, that can be best explained by a simple space diagram, similar to Fig.1a, and showing a flattened ellipsis shape of the apparatus. 

\begin{figure}[htp]
\centering
\includegraphics[totalheight=0.15\textheight]{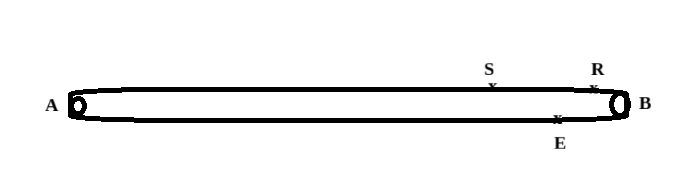}
\caption{The apparatus and the circuiting fiber}
\end{figure}

The three signs on the circuit indicate, respectively, three $\phi$-places: 

1) the place E of emission of the photon;

2) the place R of the photon's reception by the ced.

3) the place S such that the arcs $\widearc{RS}$ and $\widearc{ER}$ have the same length.

The diagram assumes, in agreement with the data of the problem, that emission (resp. reception) occurs in the lower (resp. upper) section of the apparatus. To make the diagram more expressive the speed of the ced has been exaggerated.

Clearly the photon, leaving the ced at E and meeting it again at R, will not traverse the arc $\widearc{ER}$ (that is, the {\sl shorter} arc from E to R) if we think of it as a {\sl fixed arc in the laboratory-fixed apparatus}. However, since the fiber is rolling, in the time needed for the photon to reach $R$, the {\sl fiber} arc $\widearc{ER}$ will occupy the same-length {\sl fixed} arc  $\widearc{RS}$: 

\[ L(\widearc{RS}) = L(\widearc{ER}). \]

\n
Obviously, the photon, in order to reach $R$, will have to traverse the {\sl fiber} arc  $\widearc{RS}$. Thus, even if in the fixed space the photon travel covers a length $2\ell - L(\widearc{ER})$, which is lesser than $2\ell$, it traverses, nonetheless, the whole length $2\ell$ of the fiber.

This is not surprising, and it is linked to our intuitive idea of `conservation of matter'. However, the Challenge claims that, apparently, we should renounce this insight once we adopt SR. 

We shall see in detail that this claim is groundless, but that this is the foregone conclusion should be clear from re-reading our commentary to Fig. 4 with SR in mind, and by interpreting $\phi$ as a Minkowskian cs. Nothing need be changed: everything we have said remains literally true!

In order to show that nothing changes when we turn to SR, let us first prove formally, {\sl in classical kinematics}, that the counter-propagating traverses the whole fiber in a setting like that of Fig. 3 in \S 4.2. By selecting event $r$ as the space-time origin for both Galilean systems comoving with the ced we have, at rest with the lower section of the fiber, a cs we denote by $\phi''_G$, and its transition function with respect to $\phi$ is:

\be \phi''_G\circ\phi^{-1}  \; \left\{\ba{rcl} x'' &=& x - vt - x^\ast
\\ [4pt] t''&=& t- \frac{\ell - x^\ast}{v} \ea \right. \ee{gali-lower} 

\n
while at rest with the upper section we have a cs we denote by $\phi'_G$:

\be \phi'_G \circ \phi^{-1}  \; \left\{\ba{rcl} x' &=& x + vt - (2\ell - x^\ast)
\\ [4pt] t' &=& t- \frac{\ell - x^\ast}{v} \ea \right. . \ee{gali-upper}  

Since we already know the $\phi$-coordinates of the events $p^\ast, p^-, q^-$ from \rf{cophi} and \rf{qu-meno}, in order to compute what is the interval of fiber traversed in the lower and, respectively, the upper section, we only need to compute $\De_G x'' (p^\ast, p^-))$ and $\De_G x' (p^-, q^-)$. By substitution we obtain:

\[ \De_G x'' (p^\ast, p^-)= - (1+\bt) x^\ast, \; \De_G x' (p^-, q^-) = 2 \ell - (1+\bt) x^\ast, \]

\n
and therefore the {\sl total traversed fiber} (TTF) is:

\be \mbox{TTF} = |\De_G x'' (p^\ast, p^-)| + |\De_G x' (p^-, q^-)| = 2 \ell. \ee{totalf-G}

\n
This proves that in classical kinematics the photon reaches back the ced after having traversed the whole fiber.

A different way to reach the same result is to use the formula:

\be \mbox{TTF} = c'' \De_G t'' (p^\ast, p^-) + c' \De_G t' (p^-, q^-), \ee{totalf-G-1}

\n
where $c''$ and $c'$ are the (constant) absolute values of the speed in the lower and, respectively, upper section of the apparatus. Now, an easy computation gives:

\be c'' = c'= c(1+\bt) \ee{twov}

\n
and, on the other hand, by \rf{qu-meno}:

\[ \De_G t'' (p^\ast, p^-) + \De_G t' (p^-, q^-)= t (q^-) - t(p^\ast) = t (q^-) = \frac{2\ell}{c(1+\bt)}. \]

\n
Clearly, by inserting the values given in \rf{twov} into \rf{totalf-G-1}, we again obtain \rf{totalf-G}:

\be \mbox{TTF} = c'' \De_G t'' (p^\ast, p^-) + c' \De_G t' (p^-, q^-) = 2\ell. \ee{totalf-G-2}

Notice that if we substitute $p^-$ with {\sl any other event}, the previous equation would be still valid, and therefore also \rf{totalf-G-2}. We shall see that a very different situation holds in SR.

In the next section we shall see that {\sl both procedures} lead to the correct equality also in SR.

\section{Introducing Lorentz transformations}

In SR we use $\phi''$ to denote the Minkowskian cs at rest with the fiber in the lower section of the apparatus, and $\phi'$ for the Minkowskian cs at rest with the fiber in the upper section of the apparatus. We adopt the suggestion of setting the experiment so that the counter-propagating photon reaches $A$ at the same $\phi''$-time that the ced reaches $B$.\footnote{Actually in \cite{sp25} it is assumed that in the lower section the ced moves leftwards and the photon rightwards, but this is irrelevant.}

\begin{figure}[htp]
\centering
\includegraphics[totalheight=0.4\textheight]{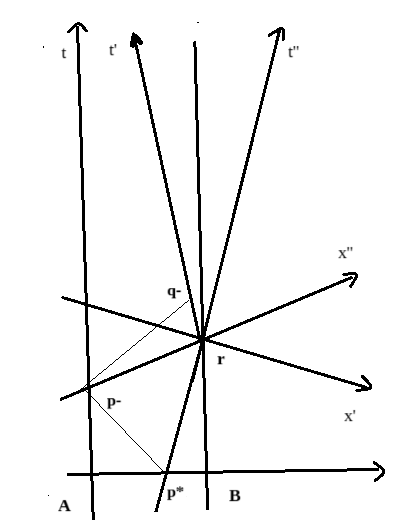}
\caption{Fixing $p^\ast$}
\end{figure}

By choosing, as in the classical approach discussed in the previous section, the space-time origin for both $\phi''$ and $\phi'$ at the turning event $r$ (cf. \rf{cophi}), we have that the Lorentz transformations from $\phi$ to $\phi''$ and from $\phi$ to $\phi'$ are, respectively:

\be \phi''\circ \phi^{-1} \; \left\{\ba{rcl} x'' &=& \al (x-\ell-v (t-\dss\frac{\ell- x^\ast}{v})) \\ [6pt] t'' &=& \dss\al ( t- \frac{\ell-x^\ast}{v} - \frac{v (x-\ell)}{c^2}))\ea\right.
\ee{LT-lower}

\vs

\be \phi'\circ \phi^{-1}\; \left\{\ba{rcl} x' &=& \al  (x-\ell +\dss v (t-\frac{\ell- x^\ast}{v})) 
\\  [6pt] t' &=& \al (t - \dss\frac{\ell-x^\ast}{v} +\dss\frac{v (x-\ell)}{c^2}) \ea\right.
\ee{LT-upper}

By requiring that $t'' (p^-) = 0$ we get 

\[ 0 = \frac{\al (1+\bt)}{\bt c} (x^\ast - (1-\bt) \ell), \]

\n
and therefore \( x^\ast = (1-\bt)\ell\). By substitution in \rf{cophi} and \rf{qu-meno} we get:

\be \ba{rcl} p^\ast &\equiv& ((1-\bt)\ell, 0), \\  [4pt] p^- &\equiv& (0, \dss (1-\bt)\frac{\ell}{c}) \\ [4pt] r &\equiv& (\ell, \dss\frac{\ell}{c}) \\  [4pt] q^- &\equiv&  (\dss\frac{1+\bt^2}{1+\bt}\ell, \frac{2\ell}{c(1+\bt)}). \ea\ee{qu-stella}

Notice that the constraint on the value of $x(p^\ast)$ fixes also the value of $x(q^-)$. Neither can possibly be equal to $\ell$.

By inserting the $\phi$-coordinates of $p^\ast$, and taking into account the change from $v$ to $-v$, equations \rf{LT-lower} and \rf{LT-upper} simplify and become, respectively: 

\be \left\{\ba{rcl} x'' &=& \al (x-v t- (1-\bt)\ell) \\ t'' &=& \al (t- \dss\frac{vx}{c^2} - (1-\bt)\frac{\ell}{c}) \ea\right.
\ee{LT-lower-1}

\be \left\{\ba{rcl} x' &=& \al (x +vt - (1+\bt)\ell) \\  t' &=& \al (t + \dss\frac{vx}{c^2} - (1+\bt)\frac{\ell}{c})\ea\right.
\ee{LT-upper-1}

Simple computations give, for the photon's worldline from $p^\ast$ to $p^-$:

\[ \ba{rcl}  p^\ast  &\stackrel{\phi''}{\equiv}&  ( 0, -\frac{\al \ell}{c}(1-\bt^2)). 
\\ [ 6pt] p^- &\stackrel{\phi''}{\equiv}&  (-\al \ell (1-\bt^2), 0) = (-\ell\sqrt{1-\bt^2}, 0), \ea \] 

\n
and for the photon's worldline from $p^-$ to $q^-$:

\be \ba{rcl} p^-  &\stackrel{\phi'}{\equiv}& \dss(- \ell \frac{1+\bt^2}{\sqrt{1-\bt^2}}, -\frac{2\ell\bt}{c\sqrt{1-\bt^2}})
\\ [6pt] q^- &\stackrel{\phi'}{\equiv}&  (0, \dss \frac{\ell}{c}\frac{(1-\bt)^2}{\sqrt{1-\bt^2}}). \ea \ee{pq-meno-phi-1}
  
As done in the classical kinematics approach, we define, with subscript $R$ instead of $G$, $\De_R x'' (p^\ast, p^-)$, and $\De_R x' (p^-, q^-)$ and by simple computations we obtain:

\[ \De_R x'' (p^\ast, p^-) = - \ell \sqrt{1-\bt^2}, \; \De_R x' (p^-, q^-)= \ell \frac{1+ \bt^2}{\sqrt{1-\bt^2}}. \]

Thus, by taking into account \rf{properlength}, the length of the fiber traversed by the photon turns out to be:

\be \mbox{TTF}_R = |\De_R x'' (p^\ast, p^-)| + |\De_R x' (p^-, q^-)| = \frac{2 \ell}{\sqrt{1-\bt^2}} = 2 \ell_0.  \ee{totalf-R}

{\sl This is the exact translation in SR of the equality \rf{totalf-G}.} 

One might ask whether even the second procedure works in the case of SR. This time we know apriori that $c'' = c'= c$. On the other hand:

\[ \ba{rcl} \De_R t'' (p^\ast, p^-) &=& \frac{\ell}{c\al},
\\ [6 pt]  \De_R t' (p^-, q^-) &=& \frac{\ell}{c}\al (1+\bt^2), \ea \]

\n
and therefore

\[ c\De_R t'' (p^\ast, p^-) + c \De_R t' (p^-, q^-) = 2\ell\al = 2 \ell_0, \]

\n
just as it should be. 

So why should the SR treatment be deemed to be perplexing?

\section{The objection}

According to \cite{sp25}, the length of the traversed fiber in SR should be not \rf{totalf-R}, but the sum of the products of $c$ with, respectively, $\De_R t''(p^\ast, r)$ (called $T_{out}$) and $\De_R (r, q^-)$ (called $T_{ret}$), that is, by \rf{proti}:

\[\ba{rcl}  cT_{out} &=& |r-p^\ast| = \dss \ell\sqrt{1-\bt^2} 
\\ [8pt] cT_{ret} &=& |q^- - r| = \dss \ell\frac{(1-\bt)^2}{\sqrt{1-\bt^2}} \ea,\]

\n
that is

\[ \ba{rcl} cT_{out} + cT_{ret} &=& \ell (\sqrt{1-\bt^2} + \dss\frac{(1-\bt)^2}{\sqrt{1-\bt^2}}) 
 = \dss\frac{2\ell (1-\bt)}{\sqrt{1-\bt^2}}
\\ [6pt] &=& 2\ell_0 - 2\ell_0 \bt, \ea \]

\n
thus

\be cT_{out} + cT_{ret} = 2\ell_0 - 2\ell_0 \bt. \ee{timegap-1}

\n
The residuum 

\[ c\de t := 2\ell_\bt \]

\n
is what has been called the ``missing section'', while 

\[ \de t' : = \al \de t  = 2\ell_0 \bt\] 

\n
which assumes the $\phi'$'s viewpoint, is the ``time gap'', interpreted as a ``discontinuity'' in space-time. 

We have seen that, by following both the right procedures, one obtains the right result, with no residuum. The reason is that we cannot obtain the correct result by using proper times and multiplying them by $c$, just because
 
 \[ cT_{out} = |\De x'' (p^\ast, p^-)|, \;\; \mbox{but}\; \; cT_{ret} < |\De x' (p^-, q^-)|, \]

\n
so there must be a residuum not accounted for by the sum \(cT_{out} + cT_{ret}\).

One might wonder: but why does that inequality come up in the case of $cT_{ret}$? 

The best and most concise answer to this seems to be to ask another question: {\sl why should we expect that a coordinate system formed by pasting together two Minkowskian systems should behave like a single Minkowskian system?} In fact this pasting together does generate not even any {\sl global} cs. So the assumption that the ced's proper time should be equivalent to the time coordinate of a cs is clearly unwarranted. 

\section{The twin paradox in disguise}

In fact, the feeling of weirdness in the asserted ``time gap'' has its source in a strictly similar uneasiness with what is known as the twin paradox.  

I will illustrate the point with direct application to the problem we are dealing with, referring to \cite{mm22} and to the Appendix for other remarks.

\begin{figure}[htp]
\centering
\includegraphics[totalheight=0.4\textheight]{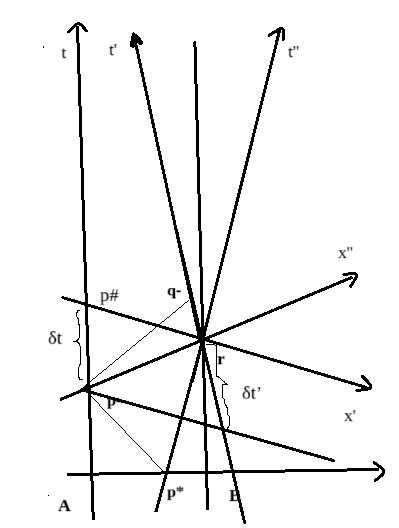}
\caption{The twin paradox behind the ``time gap''}
\end{figure}

In Fig. 6 it is clear that the segment of $A$'s worldline with endpoints the events $p^-$ and $p^\#$ disappears, so to speak, in the abrupt passage from $\phi''$ to $\phi'$. 

Let us compute $t(p^\#)$ by requiring $t'(p^\#) = 0$ and $x(p^\#) = 0$ in LT \rf{LT-upper-1}; this gives immediately

\[ t(p^\#) = \frac{\ell}{c}(1+\bt) \]

\n 
and therefore the `omitted' $\phi$-time is, by inserting \rf{cophi}$_2$:

\[ \de t: = t(p^\#) - t(p^-) = \frac{2\ell}{c}\bt, \]

\n
which, as $\phi'$-time, is affected by time dilatation:

\[ \de t': = \al \de t = \dss \frac{2\ell\bt}{c\sqrt{1-\bt^2}}. \]

With these definitions we recover exactly \rf{timegap-1}. Notice that, comparing this result with the expression of $p^-$ in $\phi'$ as given in 
\rf{pq-meno-phi-1}, we  get:

\[ \de t'= | t'(p^-)|, \]

\n
just as is represented in Fig. 6.

A treatment of the classical twin paradox pointing out that here we are essentially presented with the same phenomenon is sketched in the Appendix.

\section{Conclusion}

I have shown that the supposed ``time gap'' and ``missing section'' arise from using an inadequate formula in computing the `total traversed fiber' length in the counter-propagating photon's worldline. By using in SR the same correct procedures used in classical kinematics, one gets the right result with no residuum whatsoever. I also stressed that SR predicts correctly the LSE.

However, there is no denying that special relativity {\sl is} counter-intuitive. In the LSE it is certainly annoying to our `classical' habits that the rolling optical fiber in the apparatus can be entirely described as a single spatial entity only insofar as the apparatus is at rest in the laboratory and the motion of the fiber is assumed to be bound to it. Instead, the lower section of the fiber and the upper section of the fiber define different rest spaces, and therefore, they do not form, together, a single spatial entity in any absolute way. 

This is analogous to the case of a rotating disk, whose motion can be described relativistically in the laboratory frame's rest space, but it is not, in itself, a single spatial entity in any absolute way (cf. \S 5 of \cite{mm22}).

The history of debates on the twin paradox is evidence enough that the fact that an accelerated observer in SR does not define a global coordinate system, let alone a Minkow\-skian one, has been found disturbing by many physicists.

Now, when a debate on a scientific matter lasts for so long on a consistency issue, the most plausible explanation is, usually, that some participants are unconsciously introducing in their arguments presuppositions which are alien to the conceptual framework of the criticized theory.

This is not to put any theory beyond conceptual or logical criticism. We must never forget that the most excoriating conceptual criticism to the foundations of special relativity  came from one of its creators: Albert Einstein; and that it is to this criticism that we owe the rise of such a conceptually challenging theory - and a widely debated one in its turn, of course - as general relativity.

A personal recollection may not be out of place at this point.

My experience with logical criticisms of special relativity has made me suspicious of the chances of success by their proponents.

Since my first published paper on special relativity \cite{bm}, I have held the view that the empirical support of special relativity has to be investigated with the utmost attention, and that to assume that this support is, or may ever become, ``beyond a shadow of a doubt''  (\cite{will}, p. 245) is an improper attitude - with more of a whiff of party line: something which, in my view, should be avoided at all costs in science. Later on, I argued that the so-called conventionality of simultaneity is not a sound solution to the conceptual difficulties of the theory (contrary to the opinion of several authors in the field, both `orthodox' and `heretics'), and that in particular the latter's relationship with quantum mechanics remains, after a century, highly unsatisfactory (\cite{mm01}, \cite{mm18}, \cite{mm22}). 

However, an inconsistency in a physical theory (or between two physical theories)  which has proven to be empirically satisfactory in many settings cannot be considered a death sentence for the theory (for related remarks see also \S 11 of \cite{mm22}). It should be taken as evidence that there are parts of it that need to be put under surveillance, or even quarantined until full logical recovery (there was also a grain of epistemological, not just mundane, wisdom in the curate's reply as told in the ``Curate's Egg'' satirical story).

Be that as it may, my conclusion is that the ``Challenge'' discussed in this paper has failed to unearth a real logical trouble in special relativity, and that, basically, it is a rehearsal of `classic' perplexities concerning the twin paradox.

\section{Appendix - The twin paradox in a nutshell}

\begin{figure}[htp]
\centering
\includegraphics[totalheight=0.3\textheight]{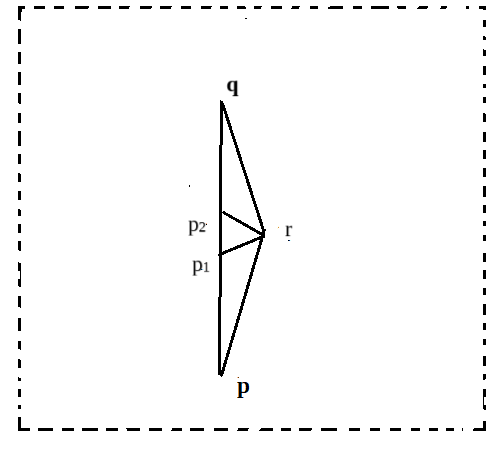}
\caption{Twin paradox}
\end{figure}

The worldlines $pq$ and $prq$ (Fig. 7) correspond to the 'sedentary' and to the `travelling' twin respectively. The paradox is that, according to the principle of proper time, the travelling twin ages less than the sedentary twin, and that they can verify it experimentally when they meet at $q$. In this appendix I will be concerned only with the geometric interpretation of the twin paradox as it bears on the ``Challenge''.

We assume, as in the picture, that 

\[ \frac{1}{c}|r-p| = \frac{1}{c}|q-r)| =: \tau\]

\n
If $|q - p| = cT$, then the surprising inequality (surprising in Euclidean, that is, not in Lorentzian, geometry) is:

\[ T>2\tau.\]

In the figure, the worldlines $pr$ and $rq$ represent motions with respect to $pq$ with speeds $v$ and $-v$ respectively ($0<v<c$).

Now, $p, p_1, p_2, q$, with respect to the timelike vector $u$ directly proportional to $q-p$ and having module $c$, can be written, by taking $p$ as the origin, as $p+ tu$, with

\[ t(p) = 0, \; t(p_1) = t_1, \; t(p_2) = t_2, \; t(q) = T. \]

\n
The events $p_1$ and $p_2$ are uniquely determined by requiring the Lorentzian orthogonality conditions:

\be g(r-p_1, r-p) = 0, \; g(r-p_2, q-r) = 0. \ee{ortho}

By the time dilatation effect, the time intervals $[0,t_1]$ and $[t_2, T]$ are both computable by the travelling twin as $\tau/\al$. Thus

\[ 2\tau = \al t_1 + \al ( T-t_2) = \al T - \al (t_2-t_1). \]

\n
A computation based on \rf{ortho} gives:
  
\[ t_2 - t_1 = 2\tau \al\bt^2 \]

\n
and by substitution in the precedent equation we obtain 

\[ 2\al^2 \tau = \al T \; \therefore \;  2\tau = T\sqrt{1-\bt^2} <T. \]

\n
Notice that the size of the ``time gap'' is such that $2\tau$ turns out to be not just lesser than $\al T$, but also lesser than $T$ - this being the  differential aging effect. 

Obviously in classical physics the events $p_1$ and $p_2$ coincide, and therefore the time difference $t_2 - t_1$ vanishes. 

\vs\vs
\n
{\bf Acknowledgment} Exchange of e-mails with Gianfranco Spavieri led me to a deeper understanding of his argument, and resulted in several improvements in the final version of the original paper.

\section{Postface}

The paper above was written, at the insistence of Gianfranco Spavieri, in order to comply with the requirements of the ``Challenge'' reproduced verbatim in \S 2. Clearly the Award's promoters expected that his claim that a serious inconsistency lurked at the basis of special relativity should have been vindicated by some of the participants.

The paper may be one of the most intensively peer-reviewed in the field. In fact I had with Spavieri and, partly through him, with the Research Group promoting the Award and assessing the submissions, a correspondence stretching to nearly three months.

Clearly, if some entity launches a competition to prove one of its cherished opinions, it strains credulity to imagine that the winner will be someone showing the falsity of that opinion. So I was not surprised by the unsuccessful outcome of the submission process, the more so as I had recommended Spavieri to ``retract, or at least publish a warning'' about, his papers based on the argument I have refuted. 

In any case, I decided to publish this paper both as my own warning to scholars in this field, and as a useful supplement to my ``Selleri'' paper \cite{mm22}. As such, I think, it should be of interest to the same audience as the former paper.

\end{document}